\documentclass[final,3p,times,onecolumn]{elsarticle}
\onecolumn
\usepackage{hyperref}
\usepackage{pifont}
\usepackage{color}
\usepackage{tabularx, booktabs}
\usepackage{amsmath}
\usepackage{algorithmic}
\usepackage{float}
\usepackage{graphicx}
\usepackage{subcaption}

\usepackage{xcolor}
\usepackage{multirow}
\usepackage{algorithm}
\usepackage{hyperref}
\usepackage{caption}
\usepackage{fancyhdr}
\usepackage{tabularx, booktabs} 
\usepackage{natbib}
\usepackage{hyperref}
\usepackage{color}
\usepackage{tabularx, booktabs}
\usepackage{amsmath}
\usepackage{algorithmic}
\usepackage{float}

\usepackage{xcolor}
\usepackage{multirow}
\usepackage{algorithm}
\usepackage{hyperref}
\usepackage{caption}
\usepackage{fancyhdr}
\usepackage{tabularx, booktabs} 
\usepackage{natbib}
\usepackage{graphicx}

\begin{document}

\begin{frontmatter}

\title{LSEAD: A Privacy-Preserving LLM-Based Speech Analysis Framework for Early Alzheimer’s Disease Screening}

\author{Xin Wang}
\address{Faculty of Digital Innovation, Arts \& Sciences, Saskatchewan Polytechnic, Regina SK S4S 5X1, Canada}
\author{Yingchao Huang\corref{mycorrespondingauthor}}
\ead{huangyi@saskpolytech.ca}
\address{Faculty of Digital Innovation, Arts \& Sciences, Saskatchewan Polytechnic, Regina SK S4S 5X1, Canada}
\author{Yuhan Su}
\address{School of Basic Medical Sciences, Hebei University, Baoding 071000, China}
\author{Shanshan Yao}
\address{Department of Civil \& Environmental Engineering and School of Mining \& Petroleum Engineering, University of Alberta, Edmonton AB T6G 2H5, Canada}
\author{Wei Peng}
\address{Faculty of Engineering and Applied Science, University of Regina, SK S4S 0A2, Canada}

\cortext[mycorrespondingauthor]{Corresponding author}

\begin{abstract}
Early diagnosis of Alzheimer’s disease (AD) is critical for enabling timely interventions that may slow disease progression and improve patient outcomes. However, there is a pressing need for AD detection approaches that are both non-invasive and cost-effective, particularly in real-world clinical settings where patient populations and recording conditions vary widely. Speech-based screening meets these requirements which relies on natural, low-burden data collected without specialized equipment. Recent advances in large language models (LLMs) further strengthen this approach by enabling richer linguistic representation and improved model generalization, making speech-based AD detection more robust and practical for real-world deployment. In this study, we propose an \textbf{L}LM-based \textbf{S}peech-assisted \textbf{E}arly \textbf{AD} (LSEAD) detection framework. It leverages the linguistic and clinically relevant knowledge encoded in pretrained LLMs to capture cognitive-linguistic AD markers without relying on hand-crafted features. Spontaneous patient speech is first transcribed, after which high-dimensional text embeddings are generated using locally deployed and open-source LLMs. Principal component analysis (PCA) is then applied to reduce dimensionality and extract informative features for classification. By relying exclusively on speech transcripts, LSEAD enables efficient AD risk assessment without invasive procedures or external data exchange, supporting privacy-preserving deployment. The proposed LSEAD framework is evaluated on two benchmark datasets, ADReSS20 and ADReSSo2021. Experimental results demonstrate that LLM-based embeddings generalize effectively across datasets and consistently outperform existing approaches in AD classification accuracy by 5\%, particularly for early-stage detection. These findings highlight the potential of locally deployed, LLM-driven speech analysis as a practical, secure, and scalable solution for AD screening in real-world clinical settings.
\end{abstract}

\end{frontmatter}

\section{Introduction}
\label{sec:intro}

Alzheimer’s disease (AD) is the leading cause of dementia among older adults and is characterized by the progressive deterioration of memory, language, and cognitive function \cite{kumar2024alzheimer}. It accounts for approximately 60–70\% of all dementia cases worldwide \cite{reitz2011epidemiology, who2017dementia}. Although there is currently no cure for AD, early diagnosis plays a crucial role in enabling timely interventions that may slow disease progression and improve patients’ quality of life \cite{crous-bou2017alzheimer}.

Conventional AD diagnosis relies on a combination of clinical evaluations, neuropsychological testing, and advanced neuroimaging techniques such as magnetic resonance imaging (MRI) and positron emission tomography (PET) \cite{c2bc6ca63e9047f185981d4b59b748b9, KAUR2024102928, KHOJASTESARAKHSI2022102332}. While these approaches provide valuable diagnostic and prognostic information, they are often resource-intensive and costly, require specialized equipment and trained clinical personnel, and are typically confined to hospital or specialized clinical settings. In particular, PET imaging involves exposure to radioactive tracers and is associated with substantial financial and logistical burdens, limiting its suitability for large-scale screening or frequent monitoring. MRI, although non-invasive and free of ionizing radiation, remains expensive, time-consuming, and less accessible in community or home-based contexts. Together with comprehensive neuropsychological assessments, these constraints restrict the scalability and accessibility of conventional diagnostic pathways \cite{LAZLI2026109026}. As a result, these conventional methods are not well suited for large-scale screening, frequent monitoring, or deployment in home and community environments. These limitations highlight a pressing need for AD detection approaches that are non-invasive, cost-effective, and robust to the variability encountered in real-world clinical settings, where patient characteristics and data collection conditions can differ substantially. These limitations also highlight the need for intelligent, deployable clinical systems that can support scalable screening and early-stage decision-making outside traditional healthcare settings.

In response to the limitations of conventional diagnostic approaches, recent research has increasingly focused on non-invasive and cost-effective alternatives that can be deployed outside specialized clinical settings. These methods leverage multimodal signals that reflect cognitive and neurological function while reducing patient burden and resource requirements. Representative approaches include electroencephalography (EEG), which enables real-time measurement of brain activity \cite{sudharsan2022recognition, safi2021early, TAUTAN2021102081}; eye-tracking, which captures abnormalities in gaze behavior associated with deficits in memory and attention \cite{tokushige2023early, TAUTAN2021102081}; facial expression analysis, which reflects diminished facial muscle activity and altered affective responses in individuals with AD \cite{doi:10.1176/appi.neuropsych.21070186, zheng2023detecting}; and speech analysis, which examines changes in acoustic features and linguistic characteristics that accompany cognitive decline \cite{konig2025novel}.

Among these non-invasive techniques, speech-based analysis has emerged as one of the most promising approaches for early AD detection. Speech is a natural, low-burden signal that can be collected unobtrusively without specialized equipment, making it particularly well suited for scalable, remote, and repeated screening. Importantly, speech data in clinical research are typically elicited using structured picture description tasks, in which participants are asked to describe a standardized image. This paradigm provides a controlled and reproducible elicitation protocol that minimizes variability in topic, content, and task demands across individuals \cite{becker1994natural}. Picture description tasks are widely used in neuropsychological and clinical assessments because they simultaneously engage semantic memory, lexical retrieval, syntactic planning, and discourse organization—cognitive processes known to be affected early in AD \cite{Filiou02062020, luz2020alzheimersdementiarecognitionspontaneous}. The controlled nature of these tasks ensures the comparability of speech samples across different datasets, while also allowing subtle linguistic impairments to emerge. Collectively, these properties position speech analysis based on picture description tasks as a practical, accessible, and cost-effective solution for non-invasive cognitive screening in real-world clinical and community settings.

Building on these advances, recent developments in large language models (LLMs) provide new opportunities for developing intelligent and scalable systems for speech-based AD detection. LLMs enable the extraction of rich, high-level linguistic representations that are often difficult to model using traditional hand-crafted features or shallow learning approaches \cite{AlsuhaibaniMuath2025ARoM}. Trained on extensive datasets, LLMs can perform a wide range of tasks, from text generation to complex problem-solving, which allows for richer linguistic feature representation and improved model generalization \cite{agbavor2022predicting}.

By processing spontaneous speech or transcribed dialogue, LLMs can detect linguistic anomalies and longitudinal changes that may indicate early cognitive decline, thereby improving the sensitivity and robustness of non-invasive AD risk assessment \cite{PANESAR202320}. In addition to detection, LLM-driven systems can support personalized education and guidance by drawing on large bodies of medical knowledge to deliver tailored information and recommendations, extending their utility from AD screening to practical support for patients and caregivers \cite{bioengineering12060631}. These capabilities position LLM-based speech analysis as a promising foundation for scalable and clinically meaningful cognitive health assessment.

Motivated by the need for non-invasive, cost-effective, and scalable AD screening tools, the objective of this study is to develop and evaluate LSEAD, an \textbf{L}LM-based \textbf{S}peech-assisted \textbf{E}arly \textbf{AD} detection framework. LSEAD is a privacy-preserving, speech-based system that leverages locally deployable, open-source LLMs to capture clinically relevant linguistic features associated with cognitive decline. Specifically, this work aims to (i) extract high-level semantic representations from spontaneous speech transcripts using LLMs, (ii) enhance robustness and computational efficiency through dimensionality reduction, and (iii) rigorously evaluate classification accuracy, early-stage detection capability, and cross-dataset generalization using benchmark AD speech datasets. By integrating advanced language modelling with lightweight classification strategies, this study aims to support practical clinical decision-making and enable scalable, real-world AD screening. 


\section{Related work}
\label{sec:related_work}

Human speech analysis generally involves both acoustic features, such as formants, pitch, and phonemes, and linguistic features, including morphemes, words, sentence structures, and contextual meaning \cite{SHADLE2006442}. Because speech production reflects both physical mechanisms and higher-level language processes, it provides rich cues for detecting cognitive impairment. However, acoustic features alone are not always reliable indicators of medical conditions, particularly in older adults \cite{AlsuhaibaniMuath2025ARoM}, since age-related physiological and peripheral changes can significantly alter acoustic characteristics after the age of 60, independent of cognitive decline \cite{MAHON202222}.

With recent advances in LLMs, research has increasingly focused on linguistic features for cognitive impairment detection. Owing to their strong natural language understanding capabilities, LLMs can effectively capture subtle changes in grammar, word choice, and discourse patterns associated with AD \cite{Zolnour2025LLMCARE}. This progress is largely driven by transformer-based architectures and robust embedding techniques that enable deep neural networks to model complex language representations \cite{AlsuhaibaniMuath2025ARoM}. Although linguistic features can vary across individuals due to differences in education and life experience, deep learning approaches trained on large speech corpora have demonstrated the ability to identify consistent impairment-related patterns across diverse populations \cite{Shankar2025XAI}.

A critical step in linguistic analysis is speech transcription, particularly for datasets that do not provide manual transcripts, as transcription quality has a direct and measurable impact on downstream AD detection performance \cite{info:doi/10.2196/78082}. As a result, Automated Speech Recognition (ASR) systems are commonly employed to convert recorded speech into text, forming a foundational preprocessing stage in speech-based AD detection pipelines \cite{RUSSELL2024100163, ZHANG2025106821}. Prior work has shown that end-to-end voice-based systems can effectively leverage ASR-generated transcripts for AD assessment, even in fully automated clinical screening settings \cite{Agbavor2023AIADVoice}. Large-scale benchmarking efforts such as the ADReSS Challenge further standardized transcription and evaluation protocols, enabling fair comparison of speech-based AD detection methods and highlighting the role of spontaneous speech in revealing cognitive impairment \cite{luz2020alzheimersdementiarecognitionspontaneous}.

Once transcripts are obtained, textual data are transformed into word embeddings that encode semantic and contextual information in multi-dimensional vector spaces \cite{Asudani2023Embeddings}. Numerous studies have investigated embeddings extracted from pretrained language models such as BERT \cite{devlin2019bert} and GPT \cite{Radford2019GPT2}, often applying identical classifiers to isolate the contribution of linguistic representations to AD detection performance \cite{Searle2020InterspeechAD, Roshanzamir2021TransformerAD}. More recent work has extended this paradigm by leveraging LLMs to extract higher-level linguistic and evaluative features from transcribed speech, demonstrating improved robustness and diagnostic accuracy in spontaneous speech-based AD recognition \cite{https://doi.org/10.4218/etrij.2023-0356}. Other studies have focused on comparing different detection strategies while holding embedding representations constant, further emphasizing the central role of linguistic feature modeling in speech-based AD diagnosis \cite{Guo2021CookieTheftBERT, 10.3389/fcomp.2020.624488}.

Despite the growing promise of LLM-based approaches for analyzing transcribed speech, their integration into real-world clinical systems remains constrained by privacy and deployment challenges. Many high-performing commercial LLMs, such as ChatGPT, are closed-source and rely on cloud-based infrastructures that are not fully compliant with the Health Insurance Portability and Accountability Act (HIPAA), raising significant concerns regarding the handling of sensitive patient data \cite{Price2019PrivacyBigData, Moore2019HIPAAReview}. Even when HIPAA-compliant solutions are available, they are typically proprietary, limiting transparency, adaptability, and control over data governance \cite{Yadav2023DataPrivacyAI}. At the same time, hospital IT environments often require AI systems to be deployed locally within secure and isolated infrastructures, a requirement that cloud-dependent commercial LLMs cannot reliably satisfy \cite{Li2023AIHealthChatbots}. Collectively, these limitations pose a significant barrier to the clinical adoption of advanced LLM technologies for speech-based AD detection.

Consequently, existing approaches remain largely dependent on cloud-based or proprietary models, and relatively little attention has been given to developing fully deployable, privacy-preserving systems that can be seamlessly integrated into clinical environments while maintaining strong diagnostic performance. To address this gap, this work investigates the use of open-source, locally deployable LLMs that are explicitly compatible with clinical privacy and security requirements. Recent studies have shown that privacy-preserving frameworks built on on-premises open-source LLMs can achieve competitive diagnostic performance while enabling compliance with healthcare regulations and institutional deployment constraints \cite{Mortensen2025ADetectoLocum}. By ensuring that all speech data are processed, stored, and analyzed entirely within the healthcare provider’s infrastructure, such approaches eliminate the need for external data transmission and inherently preserve patient privacy. Building on this foundation, the proposed LSEAD framework adopts a locally deployable open-source LLM, enabling secure, transparent, and practical integration into hospital systems while maintaining strong performance for speech-based AD detection.

Among locally deployable open-source LLMs, Zephyr represents a particularly suitable candidate for privacy-preserving speech-based AD detection. Zephyr-7B-$\beta$ is a chat-optimized, open-source model derived from the mistralai/Mistral-7B-v0.1 architecture and fine-tuned on a mixture of public and synthetic datasets, enabling performance comparable to commercial models such as GPT-3.5 \cite{brown2020language}. Prior studies have demonstrated that proprietary GPT models can effectively analyze speech-derived text for AD detection; however, their closed-source and cloud-based deployment prevents full compliance with healthcare privacy regulations, including HIPAA \cite{agbavor2022predicting}.

As a beta-stage model, Zephyr-7B-$\beta$ may exhibit output variability or bias in generative applications, and its training data cutoff (generally early 2023) may not cover the most recent knowledge. These limitations are less critical in the context of the proposed framework, as Zephyr is used exclusively for embedding extraction rather than text generation or factual reasoning. In this setting, the primary requirement is the model’s ability to encode stable and discriminative linguistic representations from patient speech, rather than to generate precise or up-to-date textual outputs. The experimental results presented in this study demonstrate that Zephyr-7B-$\beta$ produces robust and consistent embeddings for AD classification, supporting its suitability for this task despite the inherent constraints associated with beta-stage models.

Crucially, unlike commercial LLMs, open-source models such as Zephyr-7B-$\beta$ can be deployed entirely within secure, on-premises clinical environments. This capability is particularly important for speech-based AD detection, as patient speech constitutes personally identifiable information that would otherwise require extensive de-identification before being transmitted to external cloud platforms. Local deployment eliminates this requirement by ensuring that data are curated, stored, and analyzed exclusively within the healthcare provider’s infrastructure, resulting in an inherently privacy-preserving solution. The development and use of Zephyr-7B-$\beta$ therefore reflect an evolving LLM landscape in which advances in model performance are increasingly aligned with ethical, regulatory, and real-world deployment considerations.

Building on this foundation, extensive experimental evaluations demonstrate that Zephyr-7B-$\beta$ delivers state-of-the-art results, exceeding the performance of established commercial models, including commercial LLMs such as GPT-3.5 and GPT-4 \cite{brown2020language}. By integrating robust ASR with Zephyr-7B-$\beta$-based feature extraction and applying principal component analysis (PCA) for dimensionality reduction, LSEAD achieves an accuracy improvement of at least 5\%. Beyond overall accuracy gains, the model exhibits strong generalization across multiple datasets, which is a critical requirement for reliable clinical deployment. Notably, it also demonstrates enhanced sensitivity to early-stage AD, enabling earlier identification of cognitive decline and supporting timely intervention to improve long-term patient outcomes. Collectively, these results indicate that the proposed framework effectively bridges the gap between advanced LLM research and practical, privacy-preserving healthcare applications, offering a viable pathway for integrating language-based AI technologies into real-world AD diagnosis.

These findings are supported by extensive experiments conducted on benchmark speech datasets, including ADReSS20 \cite{luz2020alzheimersdementiarecognitionspontaneous} and ADReSSo2021 \cite{luz2021detecting}. Across both datasets, the proposed LSEAD framework consistently outperforms existing approaches in AD classification tasks, further validating its robustness, generalizability, and suitability for real-world clinical use. All code used will be available at https://github.com/kelci2017/AD\_Text\_LLMs

\section{Materials and methodology}
\label{methodology}

\subsection{Dataset}

We evaluate the proposed LSEAD framework using the ADReSS20 and ADReSSo2021 datasets, as summarized in Table~\ref{data_table}. The ADReSS20 dataset contains 54 AD and 54 cognitively normal (CN) subjects in the training cohort, and 24 AD and 24 CN subjects in the test cohort. The ADReSSo2021 dataset includes 87 AD and 79 CN subjects in the training cohort, along with 35 AD and 36 CN subjects in the test set. Both datasets consist of spontaneous speech recordings in which participants describe the Cookie Theft picture from the Boston Diagnostic Aphasia Examination (BDAE) \cite{goodglass2001bdae}, along with accompanying clinical information such as Mini–Mental State Examination (MMSE) scores. The BDAE and MMSE are widely used clinical assessments that support the diagnosis of AD and related cognitive impairments \cite{Teipel981}. Participants in both datasets include individuals diagnosed with AD and cognitively normal (CN) controls, with training and testing cohorts stratified by age and gender to reduce demographic bias.

While the two datasets share a common task and data collection protocol, they differ in the modalities provided. The ADReSS20 dataset includes full enhanced audio recordings, normalized and sub-chunked audio segments, and manual transcripts. In contrast, the ADReSSo2021 dataset provides only full enhanced audio recordings without transcripts. Consequently, the two datasets differ substantially in temporal structure, segmentation granularity, and access to linguistic annotations. These discrepancies introduce acoustic, temporal, and linguistic distribution shifts. To ensure methodological consistency across datasets and to reflect realistic deployment scenarios, we retain only the full enhanced audio from both datasets for all experiments.

\begin{table}[]
\caption{Description of the datasets used in this study. }
\begin{center}
\begin{tabular}{c|c|c|c}
\hline
\textbf{}                          & \textbf{Subject} & \textbf{ADReSS20} & \textbf{ADReSSo2021} \\ \hline
\multirow{2}{*}{\textbf{Training}} & AD               & 54                & 87                   \\ \cline{2-4} 
                                   & CN               & 54                & 79                  \\ \hline
\multirow{2}{*}{\textbf{Test}}     & AD               & 24                & 35                   \\ \cline{2-4} 
                                   & CN               & 24                & 36                   \\ \hline
\multirow{2}{*}{\textbf{Total}}    & AD               & 78                & 122                  \\ \cline{2-4} 
                                   & CN               & 78                & 215                  \\ \hline
\end{tabular}
\end{center}
\label{data_table}
\end{table}

\subsection{Method}

\begin{figure*}[!ht]
    \centering
    \includegraphics[width=0.9\textwidth]{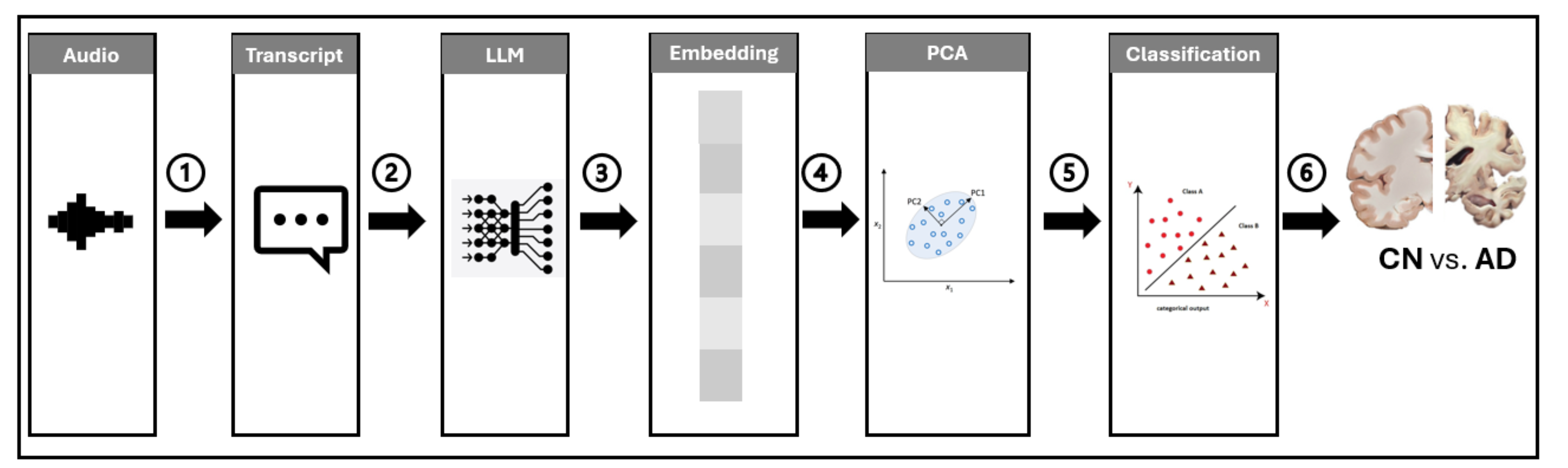}
    \caption{Overview of the proposed LSEAD framework.}
    \label{scheme_fig}
\end{figure*}

This study proposes LSEAD, a speech-based clinical screening framework designed to transform raw patient speech recordings into compact and semantically informative representations for reliable AD detection. The framework is designed as an end-to-end, deployable system that supports non-invasive and privacy-preserving cognitive assessment in real-world clinical environments. As illustrated in Figure~\ref{scheme_fig}, the system follows a sequential processing pipeline beginning with patient audio recordings. The raw speech signals are first transcribed into textual form using an ASR system. This design choice intentionally prioritizes linguistic content over acoustic features, motivated by prior studies demonstrating that textual information captures the most salient cognitive and semantic impairments associated with AD \cite{https://doi.org/10.4218/etrij.2023-0356, pan21c_interspeech}. This transcription step forms a critical component of the end-to-end clinical pipeline, enabling downstream language models to operate on standardized and interpretable textual inputs.


The resulting transcripts are then fed into an LLM, where semantic and syntactic patterns associated with cognitive decline are implicitly modelled. Rather than relying on the final prediction layer, we extract high-level feature representations from the penultimate layer of the LLM. These embeddings capture nuanced linguistic characteristics relevant to AD, leveraging the deep contextual understanding learned by the model. However, due to their inherently high dimensionality, directly using these embeddings may introduce redundancy and hinder efficient classification \cite{tsukagoshi-sasano-2025-redundancy}. To address this, PCA is applied to project the embeddings into a lower-dimensional latent space while preserving the most discriminative variance.

Finally, the reduced embeddings are provided as input to a downstream classifier, which is trained to distinguish between CN and AD participants. The classifier outputs a binary decision along with an associated confidence score, indicating the likelihood of AD presence. 

\subsubsection{Speech Signal Preprocessing}

The raw continuous-time speech signal is first preprocessed to reduce recording variability and to improve robustness for downstream automatic speech recognition. This preprocessing stage standardizes the input audio across subjects and datasets by applying a sequence of operations, including resampling, amplitude normalization, silence removal, and framing.

Given a raw speech signal \( s(t) \), the signal is initially resampled to a fixed sampling rate \( f_s \) to ensure temporal consistency across recordings:
\begin{equation}
s_r(t) = \mathcal{R}_{f_s}\bigl(s(t)\bigr),
\end{equation}
where \( \mathcal{R}_{f_s}(\cdot) \) denotes the resampling operator.

Following resampling, amplitude normalization is applied to reduce inter-speaker and inter-session variations in recording loudness:
\begin{equation}
s_n(t) = \frac{s_r(t)}{\max\bigl(|s_r(t)|\bigr)}.
\end{equation}

To suppress non-informative segments such as long pauses and background noise, non-speech regions are removed using an energy-based method or voice activity detection (VAD):
\begin{equation}
s_v(t) = \mathcal{V}\bigl(s_n(t)\bigr),
\end{equation}
where \( \mathcal{V}(\cdot) \) represents the silence removal operator.

The resulting cleaned speech signal is then segmented into overlapping short-time frames to facilitate stable processing by the ASR model:
\begin{equation}
s_f^{(k)}(t) = s_v(t)\cdot w(t - kH),
\end{equation}
where \( w(\cdot) \) is a window function, \( H \) denotes the hop size, and \( k \) indexes the frame.

Finally, the preprocessed speech representation is expressed as a sequence of framed signals:
\begin{equation}
\tilde{s}(t) = \{ s_f^{(1)}(t), s_f^{(2)}(t), \ldots, s_f^{(K)}(t) \},
\end{equation}
which serves as the standardized input to the subsequent ASR transcription stage.

\subsubsection{Automatic Speech Recognition}

The preprocessed speech signal \( \tilde{s}(t) \) is converted into a textual transcript using an ASR model:
\begin{equation}
\mathbf{T} = \mathcal{A}\bigl(\tilde{s}(t)\bigr),
\end{equation}
where \( \mathcal{A}(\cdot) \) denotes the ASR mapping from speech to text and
\(
\mathbf{T} = \{w_1, w_2, \ldots, w_N\}
\)
is the resulting word sequence.

\subsubsection{Text Embedding Extraction}

The transcript \( \mathbf{T} = \{t_1, t_2, \ldots, t_N\} \) is passed to a pretrained LLM to obtain contextualized token-level embeddings:
\begin{equation}
\mathbf{E} = \mathcal{L}(\mathbf{T}),
\end{equation}
where \( \mathcal{L}(\cdot) \) denotes the LLM encoder and
\(
\mathbf{E} = [\mathbf{h}_1, \mathbf{h}_2, \ldots, \mathbf{h}_N] \in \mathbb{R}^{N \times d}
\)
is the sequence of hidden representations from the final transformer layer, with \( d \) denoting the embedding dimensionality.

To accommodate variable-length transcripts and padded batch processing, an attention mask
\(
\mathbf{m} = \{m_1, m_2, \ldots, m_N\}, \; 
\\m_i \in \{0,1\}
\)
is used to distinguish valid tokens from padding tokens. An utterance-level embedding is then computed using attention-mask–aware mean pooling:
\begin{equation}
\mathbf{e} =
\frac{\sum_{i=1}^{N} m_i \mathbf{h}_i}{\sum_{i=1}^{N} m_i},
\end{equation}
where only non-padding token embeddings (\( m_i = 1 \)) contribute to the pooled representation.

This pooling strategy yields a fixed-dimensional utterance-level embedding
\(
\mathbf{e} \in \mathbb{R}^{d}
\),
while preventing padding artifacts and ensuring stable representations across transcripts of varying lengths. The resulting embedding captures global semantic and linguistic characteristics of spontaneous speech and serves as the input for subsequent dimensionality reduction and classification. This ensures that only valid tokens contribute to the final representation, improving robustness when processing variable-length transcripts.

\subsubsection{Dimensionality Reduction via PCA}
To improve computational efficiency and reduce redundancy in high-dimensional embeddings, PCA is employed to project the feature space into a lower-dimensional representation while preserving the most informative variance.

Let
\begin{equation}
\mathbf{X} = \{\mathbf{e}_1, \mathbf{e}_2, \ldots, \mathbf{e}_M\}^\top \in \mathbb{R}^{M \times d}
\end{equation}
denote the embedding matrix of all samples. The centred data matrix is given by
\begin{equation}
\mathbf{X}_c = \mathbf{X} - \boldsymbol{\mu},
\end{equation}
where \( \boldsymbol{\mu} \) is the mean embedding vector.

The covariance matrix is computed as
\begin{equation}
\mathbf{C} = \frac{1}{M-1} \mathbf{X}_c^\top \mathbf{X}_c.
\end{equation}

Eigenvalue decomposition is performed as
\begin{equation}
\mathbf{C}\mathbf{V} = \mathbf{V}\boldsymbol{\Lambda},
\end{equation}
where \( \boldsymbol{\Lambda} \) contains eigenvalues \( \lambda_1 \geq \cdots \geq \lambda_d \).

The number of retained principal components \( K \) is selected by variance $V_a$, which is determined by grid search \cite{liashchynskyi2019gridsearchrandomsearch} based on the suggested PCA variance $[0.9, 0.95, 0.97, 0.99, 0.999]$ \cite{chen2020pca}.

\begin{equation}
\frac{\sum_{i=1}^{K} \lambda_i}{\sum_{i=1}^{d} \lambda_i} \geq V_a.
\end{equation}

The reduced feature representation is obtained as
\begin{equation}
\mathbf{Z} = \mathbf{X}_c \mathbf{V}_K,
\end{equation}
where \( \mathbf{Z} \in \mathbb{R}^{M \times K} \).

\subsubsection{Binary Classification}

The PCA-reduced features \( \mathbf{Z} \) are used to perform binary classification between AD and CN subjects. The class label is defined as
\begin{equation}
y_i \in \{0, 1\},
\end{equation}
where \( 0 \) denotes CN and \( 1 \) denotes AD.

A supervised classifier \( f(\cdot) \) estimates the posterior probability:
\begin{equation}
\hat{p}_i = f(\mathbf{z}_i),
\end{equation}
where \( \mathbf{z}_i \in \mathbb{R}^{K} \) is the reduced feature vector.

The predicted label is given by
\begin{equation}
\hat{y}_i =
\begin{cases}
1, & \hat{p}_i \geq \tau, \\
0, & \hat{p}_i < \tau,
\end{cases}
\end{equation}
with decision threshold \( \tau = 0.5 \).

This design enables effective and efficient AD classification while maintaining flexibility in model selection, making the framework suitable for integration into real-world clinical decision support systems.

\section{Results and discussion}

To comprehensively evaluate the effectiveness of LSEAD, we conducted extensive experiments on the ADReSS20 and ADReSSo2021 datasets and systematically compared our results with representative methods reported in the literature. Beyond overall performance comparisons, we carried out a model generalization study to assess cross-dataset robustness. The experimental results and analyses presented in this section therefore provide a thorough evaluation of the proposed method in terms of classification accuracy, robustness, and generalization capability.

\subsection{Experimental setup}

To support robust model development and evaluation, we consider two complementary experimental settings based on the ADReSS20 and ADReSSo2021 datasets. These settings are designed to assess both in-distribution performance and cross-dataset generalization.

In the first setting, the two datasets are combined to form larger training and testing cohorts, thereby improving statistical robustness while preserving the original age and gender stratification defined by each challenge. In this combined setting, the training cohort comprises 141 AD and 133 CN participants, while the testing cohort comprises 59 AD and 60 CN participants, yielding 274 training samples, 119 testing samples, and 393 samples in total. This configuration allows LSEAD to learn from heterogeneous speech data collected under comparable clinical protocols, supporting a more reliable evaluation of speech-based AD detection performance.

In the second setting, we evaluate cross-dataset generalization while preserving each dataset's predefined training–testing split. The model is trained exclusively on the training split of one dataset and evaluated directly on the testing split of the other dataset, which remains strictly unseen during training, without any fine-tuning or data leakage. This experimental design explicitly assesses the robustness of the proposed framework to dataset shifts arising from differences in cohort composition, recording conditions, preprocessing pipelines, and annotation availability between ADReSS20 and ADReSSo2021.

By evaluating performance across both combined and cross-dataset settings, the experimental design provides a rigorous assessment of LSEAD’s robustness and generalization capabilities under realistic data-shift scenarios commonly encountered in clinical deployment.

\subsection{Testing results analysis}

In recent speech-based AD detection research, embeddings extracted from pretrained language models or LLMs are typically employed as input features for downstream classification \cite{Kashyap2025ExplainableDementiaLLM, eng6070163, Mo2024.08.22.24312463}. Prior work has shown that classical machine learning models, including logistic regression (LR), support vector classifier (SVC), XGBoost, and neural networks (NNs), are particularly effective when applied to such embedding representations \cite{XIAO2021102362, SHANMUGAM2022103217, BOTROS2025106997}. Motivated by these findings, we adopt the same family of classifiers to ensure methodological consistency and enable direct comparison with existing approaches in the literature.

As reported in Table~\ref{cv_test_table}, when evaluated within the complete proposed pipeline, the cross-validation results indicate that LR and SVC achieve the most stable and competitive performance, with average accuracies of 81.4\% and 80.3\% and F1 values of 81.5\% and 81\%, respectively. XGBoost also demonstrates consistent performance but remains slightly inferior to these linear models. In contrast, the NNs exhibit substantially lower cross-validation accuracy. Similar observations have been reported in prior work, which shows that NNs can underperform linear or kernel-based models during cross-validation when training data are limited or highly heterogeneous \cite{goodfellow2016deep, hastie2009elements}.

The independent test results further reinforce these observations. Among all evaluated classifiers, LR achieves the best overall performance, reaching an accuracy of 90.0\% and an F1 score of 89.7\%, demonstrating strong generalization to unseen data. This performance advantage is closely linked to the inclusion of PCA in the proposed pipeline. Although Table~\ref{cv_test_table} reports only results obtained with PCA as PCA is an integral component of the framework, additional analyses show that removing PCA and training classifiers directly on the high-dimensional LLM embeddings consistently degrades performance across all models. This degradation indicates that the raw embedding space contains redundant and noisy dimensions that obscure class-discriminative structure, which is especially detrimental in small-sample settings.

\begin{figure*}[!ht]
    \centering
    \includegraphics[width=0.7\textwidth]{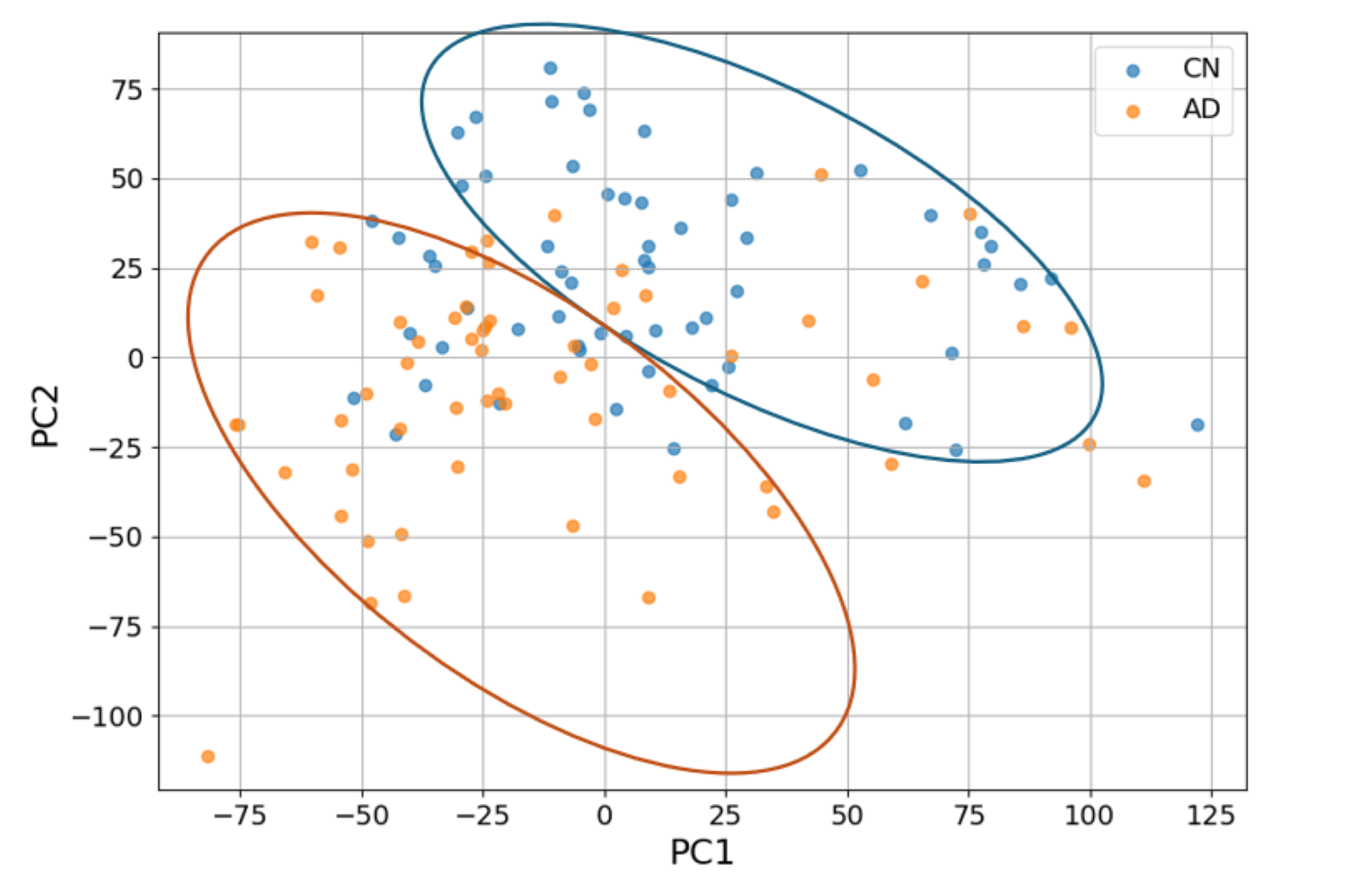}
    \caption{The PC1 and PC2 distributions for AD and CN categories.}
    \label{pca_fig}
\end{figure*}

By contrast, PCA produces a compact and denoised feature space in which the most informative variance is preserved. In this reduced-dimensional representation, linear classifiers such as LR are particularly effective, as they can exploit class-separating directions with a limited number of parameters, resulting in improved robustness and generalization. This interpretation is further supported by the PCA visualization in Figure~\ref{pca_fig}, which reveals a clear linear separation trend between CN and AD samples. CN samples are predominantly distributed in the upper-right region of the projection, whereas AD samples cluster toward the lower-left region, indicating that the dominant variance captured by the first two principal components encodes meaningful linguistic differences associated with cognitive status. These findings highlight PCA as a crucial step in the proposed framework and explain the superior performance of linear classifiers within the reduced feature space.

SVC and XGBoost also demonstrate stable and competitive performance, but they remain consistently inferior to LR under the same experimental conditions. In contrast, NNs exhibit markedly weaker performance. This outcome is expected as NNs typically require larger datasets and higher-dimensional feature representations to fully leverage their modeling capacity. Under the current setting, where the available dataset size is relatively small and dimensionality reduction has been applied to the LLM-derived embeddings, the NN model suffers from reduced parameter efficiency and increased sensitivity to information loss. As a result, it fails to outperform simpler classifiers, highlighting that appropriately chosen low-complexity models can be more effective than high-capacity architectures for small-sample AD classification tasks.

\begin{table}[]
\caption{ Model training using cross-validation and testing results. }
\begin{center}
\begin{tabular}{c|c|c|c|c|c}
\hline
                                    & \textbf{Classifier} & \textbf{Accuracy} & \textbf{Precision} & \textbf{Recall} & \textbf{F1} \\ \hline
\multirow{4}{*}{\textbf{5-fold CV}} & NNs      & 42.3\%            & 56.0\%             & 29.4\%          & 35.1\%      \\ \cline{2-6} 
                                    & SVC                 & 80.3\%            & 81.2\%             & 81.6\%          & 81.0\%      \\ \cline{2-6} 
                                    & XGBoost             & 75.9\%            & 77.7\%             & 76.0\%          & 76.3\%      \\ \cline{2-6} 
                                    & LR & 81.4\%            & 83.6\%             & 79.4\%          & 81.5\%      \\ \hline
\multirow{4}{*}{\textbf{Test}}      & NNs      & 78.2\%            & 82.4\%             & 71.2\%          & 76.4\%      \\ \cline{2-6} 
                                    & SVC                 & 82.4\%            & 86.5\%             & 76.3\%          & 81.1\%      \\ \cline{2-6} 
                                    & XGBoost             & 82.4\%            & 85.2\%             & 78.0\%          & 81.4\%      \\ \cline{2-6} 
                                    & LR & 90.0\%            & 91.2\%             & 88.1\%          & 89.7\%      \\ \hline
\end{tabular}
\end{center}
\label{cv_test_table}
\end{table}

\subsection{Early Detection Analysis}

\begin{figure*}[!ht]
    \centering
    \includegraphics[width=0.7\textwidth]{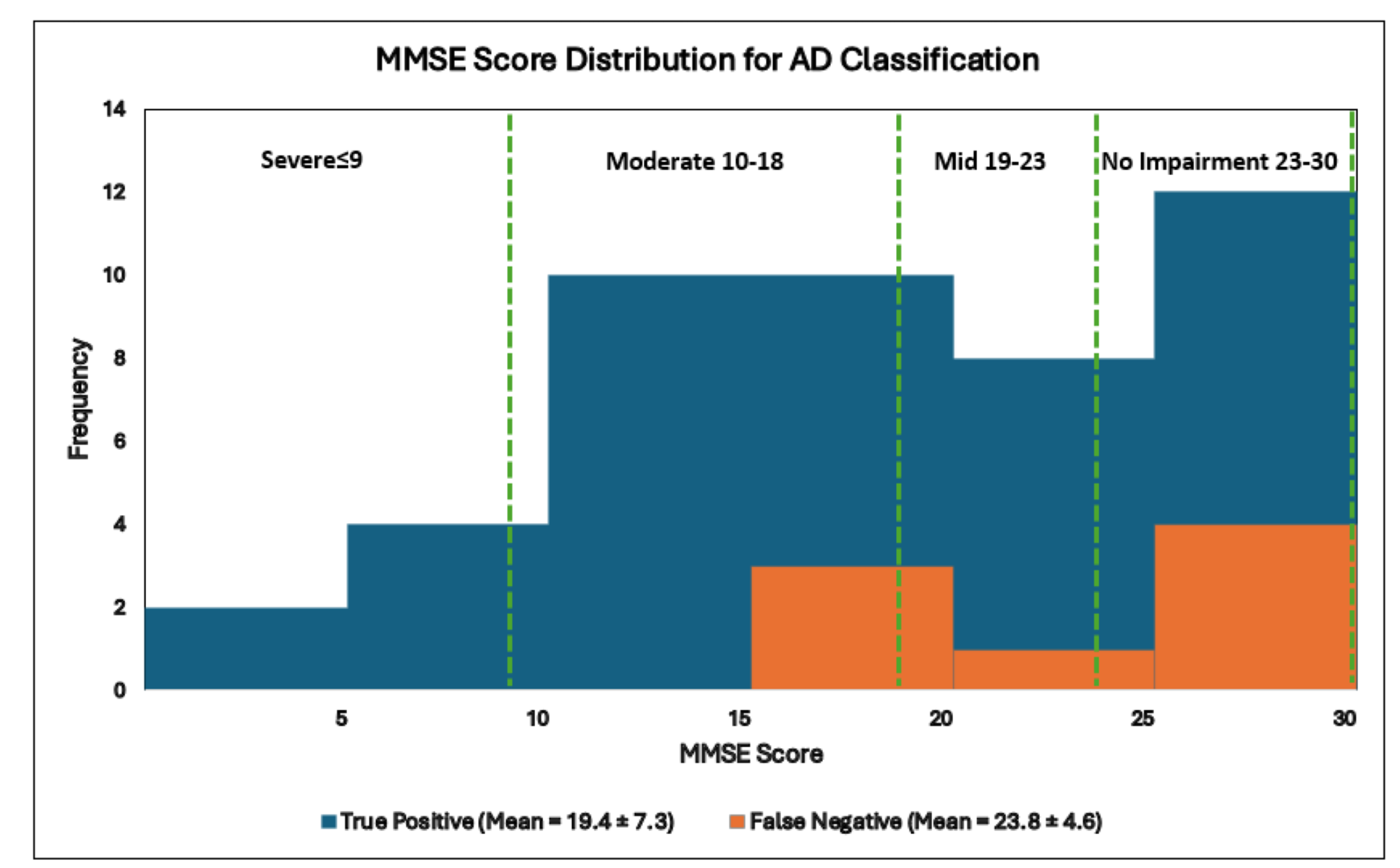}
    \caption{Histogram of correct and incorrect AD classifications with thresholds indicating MMSE cognitive impairment level. Means and standard deviations are calculated for MMSE scores per class.}
    \label{mmse_fig}
\end{figure*}

A central clinical objective of the proposed LSEAD framework is the reliable identification of AD during the early stages of cognitive decline. Figure~\ref{mmse_fig} evaluates this capability by examining model classification outcomes in relation to MMSE scores. The MMSE is a widely used cognitive screening instrument, where higher scores indicate less cognitive impairment and lower scores indicate more severe impairment \cite{Kurlowicz1999MMSE}.

Figure~\ref{mmse_fig} presents the distribution of MMSE scores for AD participants, distinguishing correctly classified AD cases (true positives) from misclassified cases (false negatives). Vertical dashed lines denote standard clinical MMSE thresholds corresponding to severe impairment ($\leq$9), moderate impairment (10--18), mild impairment (19--23), and no cognitive impairment (24--30). This stratification enables a direct assessment of the model’s performance across different stages of disease severity.

This analysis intentionally focuses on AD-positive cases because early and accurate identification of individuals with AD, particularly those with mild or borderline cognitive impairment, is the primary clinical objective of screening and early detection systems. Examining true positives and false negatives directly reveals the model’s sensitivity to early-stage disease and highlights where clinically relevant AD cases may be missed. By stratifying AD cases according to MMSE severity, this figure provides a clinically meaningful assessment of the proposed model’s effectiveness across different stages of disease progression, with particular emphasis on early and mild impairment.

As shown in the figure, correctly identified AD cases span a broad MMSE range, with a mean score of $19.4 \pm 7.3$, covering moderate and mild cognitive impairment and extending toward the early-stage region near the clinical cutoff. Notably, a substantial proportion of true positives fall within the mild impairment range (MMSE 19--23), where cognitive deficits are often subtle and more difficult to detect using conventional screening tools. This finding indicates that the proposed LSEAD framework is sensitive to early linguistic alterations associated with AD, even when global cognitive impairment remains relatively mild.

The ability to correctly classify AD cases with MMSE scores close to or above traditional diagnostic thresholds highlights the framework’s potential for early-stage detection. By leveraging linguistic representations extracted from spontaneous speech, the proposed method captures disease-related signals that may not be fully reflected in aggregate cognitive scores such as the MMSE alone. This capability is particularly important for enabling timely clinical intervention, where early identification can substantially influence disease management and treatment planning.

In contrast, false negative cases are predominantly concentrated in the higher MMSE range, with a mean score of $23.8 \pm 4.6$, corresponding largely to the transition between mild impairment and no cognitive impairment. This pattern reflects the inherent difficulty of distinguishing very early or preclinical AD, where individuals often exhibit near-normal MMSE scores and only subtle cognitive and linguistic changes. Nevertheless, the limited overlap between true positives and false negatives suggests that LSEAD maintains strong sensitivity in the early stages compared with conventional screening approaches.

The observed early detection performance is further explained by the representation-level analysis shown in Figure~\ref{pca_fig}. Although some overlap between CN and AD participants is present, reflecting the natural variability of speech patterns and the gradual progression of AD, the PCA visualization reveals noticeable differences in the centroids and principal orientations of the two distributions. This indicates that the proposed embedding framework captures systematic, disease-related linguistic differences rather than random or subject-specific variation.

Importantly, the partial overlap observed in the PCA space is consistent with the clinical reality of AD progression, where early cognitive decline may share linguistic characteristics with normal aging. Despite this challenge, the overall separation achieved in the reduced-dimensional space supports the effectiveness of the learned representations and explains the strong downstream classification performance reported in Table~\ref{cv_test_table}. Together, the MMSE-based analysis and PCA visualization demonstrate that LSEAD provides a compact, discriminative, and clinically meaningful representation of speech, enabling reliable early detection of Alzheimer’s disease in a non-invasive and scalable manner.

\subsection{Comparative results}

To provide a comprehensive evaluation of the proposed LSEAD framework, we compare its performance with several representative methods reported in the literature that were developed and evaluated on the ADReSS20 and ADReSSo2021 datasets. These comparisons position the proposed approach within the broader landscape of speech-based AD classification and enable a direct assessment of its effectiveness relative to established techniques. \par

Table~\ref{comp_table} summarizes the comparative performance of LSEAD against representative state-of-the-art speech-based AD detection methods. Across all evaluation metrics, our approach consistently achieves the strongest results, demonstrating clear advantages over existing techniques. Specifically, the proposed method attains an accuracy of 90.0\%, outperforming Mortensen et al.~\cite{mortensen2025early} by 5.1\%, Bang et al.~\cite{https://doi.org/10.4218/etrij.2023-0356} by 6.9\%, Agbavor et al.~\cite{Agbavor2023AIADVoice} by 9.7\%, and the ADReSSo challenge baseline reported by Luz et al.~\cite{luz2021detecting} by 11.1\%. These consistent improvements across diverse prior methods highlight the robustness and effectiveness of the proposed framework for speech-based AD classification.

Beyond accuracy, the proposed model achieves the highest precision (91.2\%) among all compared approaches, substantially exceeding the precision reported by Agbavor et al.~\cite{Agbavor2023AIADVoice} (72.3\%), which indicates a marked reduction in false-positive predictions. This improvement is critical for clinical applicability, as high precision reduces the risk of incorrectly flagging cognitively normal individuals. Compared with Mortensen et al.~\cite{mortensen2025early} and Bang et al.~\cite{https://doi.org/10.4218/etrij.2023-0356}
, which report balanced but lower precision and F1 scores, our method achieves a superior balance between sensitivity and specificity.

In terms of recall and F1 score, LSEAD maintains strong sensitivity (88.1\%) while achieving the highest F1 score (89.7\%), reflecting a more reliable overall classification performance than all competing methods. While Agbavor et al.~\cite{Agbavor2023AIADVoice} report high recall (97.1\%), this comes at the expense of substantially lower precision, whereas our approach provides a more clinically meaningful trade-off. Furthermore, Bang et al.~\cite{https://doi.org/10.4218/etrij.2023-0356} do not explicitly address early-stage AD detection, whereas the proposed framework demonstrates improved sensitivity to early cognitive impairment, as evidenced by MMSE-based analysis.

Compared with GPT-based approaches explored in the literature, the proposed method achieves superior performance while remaining fully locally deployable. For example, Kheirkhahzadeh et al.~\cite{Kheirkhahzadeh2023SpeechAD} reported an accuracy of only 62\% using GPT-3.5 with XGBoost, highlighting the limitations of direct API-based commercial LLM integration. In contrast, the proposed Zephyr-based framework delivers state-of-the-art results without relying on external APIs or non-HIPAA-compliant infrastructure. Comparisons with GPT-3.5 are sufficient to demonstrate effectiveness, as GPT-4 is not suitable for local, privacy-preserving deployment.

\begin{table}[]
\caption{Comparative results of our model with other contributors for AD classification. }
\begin{center}
\begin{tabular}{c|c|c|c|c}
\hline
\textbf{Contributors}          & \textbf{Accuracy} & \textbf{Precision} & \textbf{Recall} & \textbf{F1}     \\ \hline
Mortensen et al. \cite{https://doi.org/10.4218/etrij.2023-0356} & 84.9\%            & 84.7\%             & 84.7\%          & 84.7\%          \\ \hline
Bang et al. \cite{https://doi.org/10.4218/etrij.2023-0356}       & 83.1\%            & 83.1\%             & 83.1\%          & 83.1\%          \\ \hline
Agbavor et al. \cite{Agbavor2023AIADVoice}   & 80.3\%            & 72.3\%             & 97.1\%          & 82.9\%          \\ \hline
Luz et al. \cite{luz2021detecting}       & 78.9\%            & 77.8\%             & 80.0\%          & 78.9\%          \\ \hline
\textbf{Ours (LSEAD)}                  & \textbf{90.0\%}   & \textbf{91.2\%}    & \textbf{88.1\%} & \textbf{89.7\%} \\ \hline
\end{tabular}
\end{center}
\label{comp_table}
\end{table}

To highlight the effectiveness of the proposed Zephyr-7B-$\beta$ model selected for this framework, we compare its performance with two representative open-source LLMs widely adopted in medical and clinical language understanding: Meta’s Llama~2 \cite{Touvron2023LLaMA2} and Qwen3-30B \cite{yang2025qwen3technicalreport}. These models serve as strong baselines due to their demonstrated capabilities on healthcare-related benchmarks and their increasing adoption in biomedical natural language processing tasks.

Llama~2 has attracted significant attention for its strong general-purpose performance, flexibility, and open-source availability. In particular, Llama~2-7B has shown competitive results on a range of medical and clinical benchmarks, and several domain-adapted variants have been proposed for healthcare applications. Table~\ref{llama_table} reports the performance of Llama~2-7B within our pipeline when used for transcript embedding and downstream AD classification. While Llama~2 achieves competitive performance across multiple classifiers, reaching a maximum accuracy of 83.2\% with LR, it consistently underperforms the proposed Zephyr-7B-$\beta$ framework. This observation aligns with prior findings showing that Zephyr-7B-$\beta$, trained via instruction-tuned distillation, outperforms Llama~2-7B on multiple benchmarks, including those in the medical domain \cite{tunstall2023zephyrdirectdistillationlm}. In the context of speech-based AD detection, Llama~2 exhibits comparatively lower recall and F1 scores, suggesting reduced sensitivity to subtle linguistic markers associated with early-stage cognitive decline.

Table~\ref{qwen_table} presents the results obtained using Qwen3-30B, a substantially larger model that has demonstrated strong performance on medical reasoning and complex language understanding tasks. Within the proposed pipeline, Qwen3-30B yields solid classification performance, with its best results achieved using LR (87.4\% accuracy and 87.0\% F1 score). These outcomes confirm the model’s ability to extract informative linguistic representations for AD detection. However, despite its significantly larger parameter count and higher computational complexity, Qwen3-30B does not outperform the proposed Zephyr-based approach. This comparison indicates that increased model scale alone does not guarantee superior performance for speech-based AD detection, and highlights the importance of instruction tuning and task-aligned representation learning over sheer model size.

\begin{table}[]
\caption{Qwen3-30B results on AD classification. }
\begin{center}
\begin{tabular}{c|c|c|c|c}
\hline
\textbf{Classifier}          & \textbf{Accuracy} & \textbf{Precision} & \textbf{Recall} & \textbf{F1} \\ \hline
NNs      & 79.8\%            & 85.7\%             & 71.2\%          & 77.8\%      \\ \hline
SVC                 & 82.4\%            & 86.5\%             & 76.3\%          & 81.1\%      \\ \hline
XGBoost            & 80.7\%            & 83.3\%             & 76.3\%          & 79.6\%      \\ \hline
LR & 87.4\%            & 89.3\%             & 84.7\%          & 87.0\%      \\ \hline
\end{tabular}
\end{center}
\label{qwen_table}
\end{table}

\begin{table}[]
\caption{Llama 2 results on AD classification. }
\begin{center}
\begin{tabular}{c|c|c|c|c}
\hline
\textbf{Classifier}          & \textbf{Accuracy} & \textbf{Precision} & \textbf{Recall} & \textbf{F1} \\ \hline
NNs      & 73.1\%            & 86.5\%             & 54.2\%          & 66.7\%      \\ \hline
SVC                 & 79.8\%            & 81.8\%             & 76.3\%          & 78.9\%      \\ \hline
XGBoost            & 82.4\%            & 85.2\%             & 78.0\%          & 81.4\%      \\ \hline
LR & 83.2\%            & 88.2\%             & 76.3\%          & 81.8\%      \\ \hline
\end{tabular}
\end{center}
\label{llama_table}
\end{table}

\begin{figure*}[!ht]
    \centering
    \includegraphics[width=0.7\textwidth]{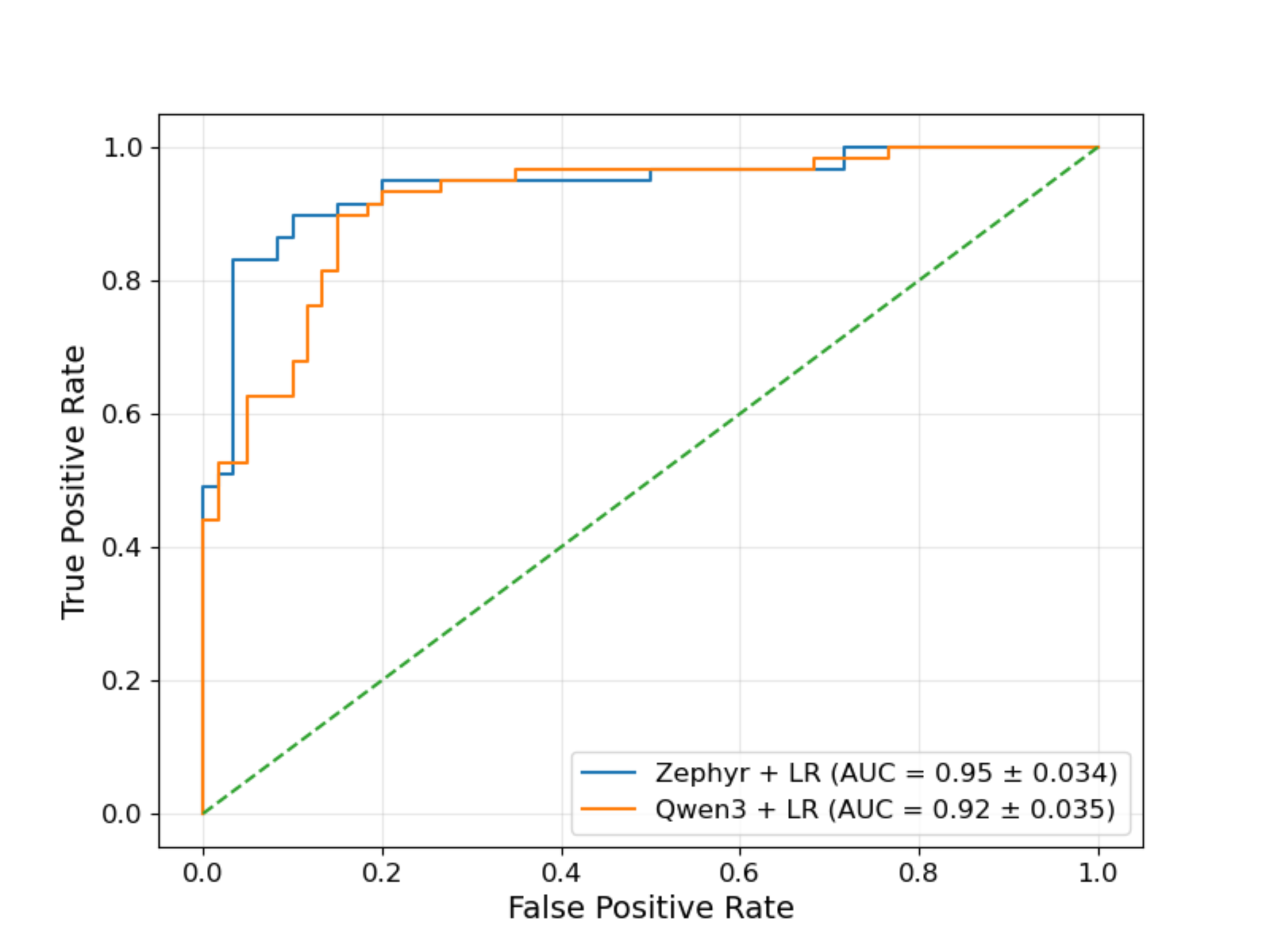}
    \caption{ROC curve comparing Zephyr-based and Qwen-based performance with 5-fold cross-validation derived AUC standard deviations.}
    \label{roc_fig}
\end{figure*}

Following the comparative evaluation of LLM backbones, Qwen3-30B demonstrates consistently stronger performance than Llama~2 across all evaluated classifiers, as shown in Tables~\ref{llama_table} and~\ref{qwen_table}. Given its superior accuracy and F1 score among the alternative LLM baselines, Qwen3-30B represents a more competitive and informative reference point for further analysis. To more rigorously assess the discriminative capability of the proposed Zephyr-7B-$\beta$ framework against a strong large-scale baseline, we therefore compare their receiver operating characteristic (ROC) curves, as illustrated in Figure~\ref{roc_fig}.

ROC analysis is widely adopted in medical decision-making and diagnostic evaluation because it provides a threshold-independent assessment of classification performance by jointly characterizing sensitivity and specificity across all operating points \cite{fawcett2006introduction, Walther2023ImagingAppropriateness}. This property is particularly important for AD detection, where the relative clinical costs of false positives and false negatives may vary depending on the screening or diagnostic context, disease stage, and downstream intervention strategy \cite{Wolfsgruber2019SCDbiomarkers}. Unlike single-point metrics such as accuracy or F1 score, ROC curves enable a more comprehensive comparison of models under varying decision thresholds, which is especially relevant for datasets with class imbalance and heterogeneous disease progression \cite{fawcett2006introduction}.

As shown in Figure~\ref{roc_fig}, the ROC curves compare the proposed Zephyr-based framework with the Qwen3 baseline for AD classification. The diagonal dashed line represents random-chance performance and serves as a reference for evaluating discriminative ability.

The proposed Zephyr model achieves a high area under the ROC curve (AUC) of $0.95 \pm 0.034$, indicating excellent ability to distinguish between AD and CN subjects. Across a wide range of false-positive rates, the Zephyr-based framework consistently achieves higher true-positive rates than the baseline model. This advantage is particularly pronounced in the low-false-positive region, which is clinically important for minimizing unnecessary follow-up assessments and reducing patient burden in screening scenarios.

In comparison, the Qwen3 baseline achieves an AUC of $0.92 \pm 0.035$, reflecting strong but comparatively lower discriminative performance. The consistent dominance of the Zephyr-based ROC curve across most operating regions demonstrates more favorable sensitivity-specificity trade-offs, suggesting that the linguistic representations extracted by Zephyr-7B-$\beta$ are more informative for AD-related speech characteristics. Furthermore, the tighter confidence interval of the proposed method indicates more stable and reliable performance across evaluation folds.

\subsection{ASR study}

To systematically examine the impact of ASR quality on downstream AD classification, we evaluated two widely adopted and empirically validated ASR systems, Wav2Vec 2.0 and Whisper, while keeping all other components of the pipeline, including the LLM, PCA configuration, and classifiers, fixed. These two ASR models were selected for complementary reasons. Wav2Vec 2.0 is a self-supervised ASR framework that has demonstrated strong performance across diverse speech tasks, particularly in low-resource and clinical settings, by learning robust speech representations directly from raw audio without reliance on large labelled datasets \cite{baevski2020wav2vec}. In contrast, Whisper is a large-scale, weakly supervised ASR model trained on massive multilingual and multitask data, and has been shown to exhibit exceptional robustness to speaker variability, background noise, accents, and recording conditions commonly encountered in spontaneous clinical speech \cite{radford2023whisper}.

The motivation for investigating multiple ASR systems is twofold. First, prior studies have shown that ASR errors are not uniformly distributed across speakers and may disproportionately affect cognitively impaired speech, potentially introducing bias or instability into downstream clinical predictions when relying on a single transcription system \cite{info:doi/10.2196/54655}. Second, in real-world clinical deployment, ASR systems may vary across institutions, evolve over time, or be replaced due to practical constraints. Evaluating multiple ASRs, therefore, provides an important assessment of the robustness and reproducibility of LSEAD under realistic operational conditions.

This analysis is particularly critical for text-based AD detection pipelines, where transcription quality directly determines the linguistic information available to subsequent language models. Even minor transcription differences can propagate through the pipeline and influence the extracted embeddings and final classification outcomes \cite{baevski2020wav2vec, radford2023whisper}. Demonstrating stable performance across different ASR systems is thus essential for ensuring clinical reliability.

As shown in Table~\ref{asr_table}, both Wav2Vec~2.0 and Whisper yield strong performance across all evaluated classifiers, confirming that the proposed framework is not tightly coupled to a specific ASR system. With Wav2Vec~2.0 transcriptions, LR achieves 88.2\% accuracy and 87.7\% F1, outperforming SVC, XGBoost, and the NNs. This result is consistent with prior findings that self-supervised ASR models produce linguistically coherent transcripts that are well suited for downstream clinical language analysis \cite{baevski2020wav2vec}.

Whisper-based transcriptions further improve overall performance, particularly for linear classifiers. Using Whisper, LR achieves the best results among all ASR-classifier combinations, with an accuracy of 90.0\% and an F1 score of 89.7\%, while SVC also benefits, reaching an accuracy of 87.4\%. These gains are likely attributable to Whisper’s robustness to speaker variability, background noise, and heterogeneous recording conditions, which are common in spontaneous clinical speech datasets \cite{radford2023whisper}. Improved transcription robustness at the ASR stage helps preserve subtle linguistic markers associated with early cognitive impairment, thereby enabling more effective embedding extraction and downstream classification.

\begin{table}[]
\caption{The test performance comparisons with different ASRs using the same LLM and setups. }
\begin{center}
\begin{tabular}{c|c|c|c|c|c}
\hline
ASR                                  & \textbf{Classifier} & \textbf{Accuracy} & \textbf{Precision} & \textbf{Recall} & \textbf{F1} \\ \hline
\multirow{4}{*}{\textbf{Wav2Vec2.0}} & NNs      & 84.00\%           & 88.50\%            & 78.00\%         & 82.90\%     \\ \cline{2-6} 
                                     & SVC                 & 82.40\%           & 82.80\%            & 81.40\%         & 82.10\%     \\ \cline{2-6} 
                                     & XGBoost             & 82.40\%           & 86.50\%            & 76.30\%         & 81.10\%     \\ \cline{2-6} 
                                     & LR & 88.20\%           & 90.90\%            & 84.70\%         & 87.70\%     \\ \hline
\multirow{4}{*}{\textbf{Whisper}}    & NNs     & 78.2\%            & 82.4\%             & 71.2\%          & 76.4\%      \\ \cline{2-6} 
                                     & SVC                 & 87.4\%            & 89.3\%             & 84.7\%          & 87.0\%      \\ \cline{2-6} 
                                     & XGBoost             & 82.4\%            & 85.2\%             & 78.0\%          & 81.4\%      \\ \cline{2-6} 
                                     & LR & 90.0\%            & 91.2\%             & 88.1\%          & 89.7\%      \\ \hline
\end{tabular}
\end{center}
\label{asr_table}
\end{table}

\subsection{Model generalization discussion}

To evaluate the generalization capability of the proposed LSEAD framework across datasets, we conducted cross-dataset validation experiments by training the model on one dataset and testing it on the other, as summarized in Tables~\ref{gen_table1} and \ref{gen_table2}. This setting is more challenging than within-dataset evaluation, as it assesses robustness to variations in recording conditions, participant demographics, and data collection protocols.

When trained on ADReSS20 and tested on ADReSSo21 (Table~\ref{gen_table1}), LR achieves the best overall performance, with an accuracy of 87.4\% and an F1 score of 86.7\%. SVC also performs strongly, yielding an accuracy of 84.9\% and a high precision of 91.8\%, indicating effective separation between AD and CN despite the domain shift. XGBoost remains competitive but shows a modest drop in performance, while the NNs exhibit lower recall, suggesting limited robustness when transferred across datasets.

A consistent trend is observed in the reverse setting, where the model is trained on ADReSSo21 and tested on ADReSS20 (Table~\ref{gen_table2}). LR again demonstrates the highest and most balanced performance, achieving 87.3\% accuracy and an F1 score of 86.6\%. SVC maintains stable performance with an accuracy of 83.1\%, whereas XGBoost and the NNs show larger discrepancies between precision and recall, reflecting increased sensitivity to cross-dataset variability.

Across both cross-dataset experiments, the proposed method exhibits stable and comparable performance regardless of the training–testing direction, highlighting its strong generalization ability. The consistent superiority of LR further supports the effectiveness of the PCA-reduced LLM embeddings, as linear classifiers are well-suited to capturing discriminative patterns in low-dimensional feature spaces. These results demonstrate that the proposed framework is robust to dataset shifts and can reliably generalize across independent clinical speech datasets for AD detection.

\begin{table}[]
\caption{Generalization results with our model trained on ADReSS20 data and tested on ADReSSo21 data. }
\begin{center}
\begin{tabular}{c|c|c|c|c}
\hline
\textbf{Classifier}          & \textbf{Accuracy} & \textbf{Precision} & \textbf{Recall} & \textbf{F1} \\ \hline
NNs      & 78.2\%            & 85.1\%             & 67.8\%          & 75.5\%      \\ \hline
SVC                 & 84.9\%            & 91.8\%             & 76.3\%          & 83.3\%      \\ \hline
XGBoost            & 82.4\%            & 83.9\%             & 79.7\%          & 81.7\%      \\ \hline
LR & 87.4\%            & 90.7\%             & 83.1\%          & 86.7\%      \\ \hline
\end{tabular}
\end{center}
\label{gen_table1}
\end{table}

\begin{table}[]
\caption{Generalization results with our model trained on ADReSSo21 data and tested on ADReSS20 data. }
\begin{center}
\begin{tabular}{c|c|c|c|c}
\hline
\textbf{Classifier}          & \textbf{Accuracy} & \textbf{Precision} & \textbf{Recall} & \textbf{F1} \\ \hline
NNs      & 81.7\%            & 95.8\%             & 65.7\%          & 78.0\%      \\ \hline
SVC                 & 83.1\%            & 84.8\%             & 80.0\%          & 82.4\%      \\ \hline
XGBoost            & 80.3\%            & 92.0\%             & 65.7\%          & 76.7\%      \\ \hline
LR & 87.3\%            & 90.6\%             & 82.9\%          & 86.6\%      \\ \hline
\end{tabular}
\end{center}
\label{gen_table2}
\end{table}

\section{Conclusions and future work}
\label{conclusion}

This study presents LSEAD, a privacy-preserving and deployable framework for speech-based AD screening that integrates LLM-derived textual embeddings with dimensionality reduction and lightweight classification. Through extensive experiments under multiple evaluation settings, the proposed system demonstrates strong performance, robustness, and generalization capability, highlighting its potential as a practical tool for non-invasive cognitive assessment in real-world clinical environments.

By leveraging pretrained LLMs, LSEAD captures subtle cognitive-linguistic patterns associated with AD that are difficult to represent using traditional hand-crafted or shallow features. The incorporation of PCA plays a critical role in improving system stability by reducing feature redundancy, suppressing noise, and producing compact representations that are well suited for small-sample clinical datasets. This design enables efficient learning while maintaining strong discriminative capability.

Across all evaluated classifiers, LR consistently achieves the best performance in both cross-validation and independent testing scenarios. This result aligns with the reduced-dimensional feature space, where linear decision boundaries are sufficient to separate cognitively normal and AD speech patterns. While SVC and XGBoost also demonstrate competitive performance, they remain slightly inferior under the same conditions. In contrast, NNs exhibit reduced effectiveness, likely due to overfitting and limited data availability. These findings suggest that, within clinically realistic data constraints, carefully designed low-complexity models can provide reliable and robust performance.

The proposed framework further demonstrates strong cross-dataset generalization, with only modest performance degradation observed when training and testing across ADReSS20 and ADReSSo2021 datasets. This indicates that the learned representations capture stable disease-related linguistic characteristics that are resilient to variations in cohort composition and recording conditions. In addition, the evaluation of multiple ASR systems shows that LSEAD maintains consistent performance across different transcription models, reinforcing its robustness and suitability for deployment in heterogeneous clinical environments.

Despite these promising results, several limitations should be acknowledged. First, the study relies on relatively small benchmark datasets, which may not fully capture the diversity of real-world patient populations, including variations in language, dialect, and comorbid conditions. Second, the current framework focuses exclusively on textual representations derived from speech, without incorporating complementary acoustic features that may provide additional diagnostic information. Third, the proposed framework currently lacks interpretability, as the decision-making process of the LLM-derived embeddings and downstream classifiers is not explicitly transparent, which may limit clinical trust and adoption.

Future work will address these limitations by extending the framework to larger and more diverse datasets, including multilingual and spontaneous conversational speech, to further validate its generalizability. In addition, multimodal approaches that integrate acoustic and linguistic features will be investigated to enhance diagnostic performance. Finally, improving interpretability remains a key priority for clinical adoption; future research will focus on explainable AI techniques to identify the linguistic markers driving model predictions, thereby enhancing transparency, supporting clinician trust, and facilitating integration into clinical decision-making workflows.


\section*{Author Contributions}

Xin Wang: Conceptualization, methodology, data curation, formal analysis, visualization, writing the original draft. 
Yingchao Huang: Conceptualization, methodology, validation, data curation, formal analysis, visualization, review and editing. 
Yuhan Su: Writing the original draft, validation, formal analysis, and visualization. 
Wei Peng: Formal analysis, visualization, review and editing. 
Shanshan Yao: Validation, visualization, review and editing.

\section*{Consent for publication}
All authors declare consent for publication.

\section*{Funding}

The authors declare that no specific funding was received for this work.



\section*{Declaration of competing interest}
The authors declare that they have no known competing financial interests or personal relationships that could have appeared to influence the work reported in this paper.

\section*{Ethics Statement}

This study involves secondary analysis of fully de-identified, publicly available datasets inclduign ADReSS20 and ADReSSo2021. No new data were collected, and no direct interaction with human participants occurred. Ethical approval and informed consent were obtained by the original data collectors, as documented in the respective dataset publications. According to institutional and national research ethics guidelines, additional ethical approval was not required for this study.

\section*{Acknowledgements}
The data used in this study were obtained from the Alzheimer's Disease ADReSS20 and ADReSSo2021 databases. We gratefully acknowledge access to these valuable open-source datasets and the support provided by Saskatchewan Polytechnic. X.Wang and Y.Huang contributed equally to this work.

\bibliography{dementia_da}

\end{document}